\documentclass[letterpaper]{article}
\usepackage{emulateapj}
\usepackage{onecolfloat}
\usepackage{apjfonts}
\usepackage{natbib}
\usepackage{graphicx}
\usepackage{amsmath}

\newcommand{\chandra}{\textit{Chandra}}
\newcommand{\Grs}	{\mbox{\rm\,GRS~1758--258}}
\newcommand{\onee}	{\mbox{\rm\,1E~1740.7--2942}}
\newcommand{\cyg}	{Cyg~X-1}
\newcommand{\xray}	{\mbox{X-ray}}
\newcommand{\xrays}	{\mbox{X-rays}}
\newcommand{\rxte}{\textit{ RXTE}}
\newcommand{\asec}	{\mbox{$^{\prime \prime}$}}
\newcommand{\amin}	{\mbox{$^{\prime}$}}
\newcommand{\aprx}	{\mbox{$\sim$}}
\newcommand{\lumin}     {\mbox{$\rm\,ergs\,s^{-1}$}}
\newcommand{\percmsq}	{$\rm cm^{-2}$}
\newcommand{\degree}	{\hbox{$^\circ$}}

\newcommand{\grsra}	{18$^h$ 01$^m$ 12$^s$.39}
\newcommand{\grsdec}	{$-$25\degree\ 44\amin\ 36\asec.1}
\newcommand{\alphara}	{18$^h$ 01$^m$ 06$^s$.1}
\newcommand{\alphadec}	{$-$25\degree\ 39\amin\ 03\asec.0}
\newcommand{\betara}	{18$^h$ 00$^m$ 31$^s$.0}
\newcommand{\betadec}	{$-$25\degree\ 41\amin\ 53\asec.9}

\begin{document}

\twocolumn[
\title{\chandra\ High Resolution Camera Imaging of \Grs} 
\author{W.A. Heindl}
\affil{Center for Astrophysics and Space Sciences, Code 0424, University of
California, San Diego, La Jolla, CA 92093}
\author{D.M. Smith}
\affil{Space Sciences Laboratory, University of California,
Berkeley, Centennial at Grizzly Peak Boulevard, Berkeley,
CA 94720-7450}


\begin{abstract}
We observed the ``micro-quasar'' \Grs\
four times with \chandra.  Two HRC-I observations were made in
2000 September-October spanning an intermediate-to-hard spectral
transition (identified with \rxte).  Another HRC-I and an ACIS/HETG
observation were made in 2001 March following a hard-to-soft
transition to a very low flux state.  Based on the three HRC images and the
HETG zero order image, the accurate position (J2000) of the
\xray\ source is RA $=$ \grsra,
Dec $=$ \grsdec\ (90\% confidence radius $=$ 0\asec.45),
consistent with the purported variable radio counterpart.  All three
HRC images are consistent with \Grs\ being a point source, indicating
that any bright jet is less than \aprx1\,light-month in projected
length, assuming a distance of 8.5\,kpc.

\end{abstract}


]  

\section{Introduction}

\Grs\ and its sister source, \onee, were the first objects dubbed
``micro-quasars'' \citep{Mir92,Mir94}.  Their \xray\ spectra are
typical of Galactic black hole candidates (BHCs), and they are
apparently associated with the time variable cores of arcminute scale
double-lobed radio sources, reminiscent of extra-Galactic radio
sources.  This morphology, seen on a parsec scale within the Milky
Way, earned them their nickname. \Grs\ and \onee\ are the brightest
persistent sources in the Galactic bulge above $\sim$50~keV
\citep{Su91}.  Their timing characteristics are typical of the black
hole low/hard state \citep{Mai99,Smi97,Hei93,Su91}, and they
consistently emit near their brightest observed levels, although
they vary over times of days to years.  Their \xray\ emission
properties are readily likened to the canonical BHC, \cyg.  In fact,
together with \cyg, they are the only known persistent, low-state
BHCs, and all three sources have maximum luminosities around $3 \times
10^{37}$ergs s$^{-1}$. Radio jets have now been observed in
\cyg, furthering the similarity \citep{Fen00}.

\Grs\ and \onee\ are, however, quite different from the Galactic {\em
superluminal} radio sources (e.g. GRS~1915+105 and GRO~J1655-40) more
typically thought of as micro-quasars.  The \xray\ emission from these
objects is much brighter and more spectacularly variable.  Their radio
jets, too, are brighter and are highly variable, being unresolved or
absent except during exceptional ejection events which last only
weeks. In contrast, the radio lobes of \Grs\ and \onee\ are quite
stable \citep{Marti02}.  

The association of \Grs\ and \onee\ with their corresponding radio
counterparts has been based primarily on relatively coarse
(\aprx10\,\asec) \xray\ positions and the very unusual (and yet nearly
identical) nature of the radio sources.  Some hint of correlated radio
and hard X-ray emission was found for \onee\ \citep{Mirabel93}, and
recently \citet{Cui01} confirmed the \xray/radio association for
\onee\ by using \chandra\ to obtain a precise \xray\ position.  In
this \emph{Letter} we do the same for \Grs, and using the fine
resolution of the High Resolution Camera (HRC) place limits on
emission from any arcsecond scale jets.
\begin{figure}
%
\plotone{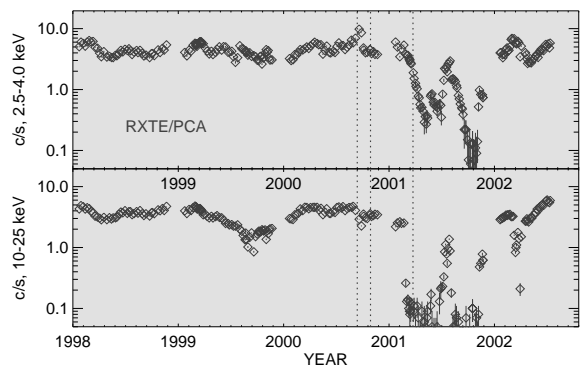}
\caption{\label{f:lc}The \rxte/PCA light curve of \Grs\ in two energy bands.
Our \chandra\ observations (see Table~1) are indicated by dashed
vertical lines.  Observations 400163 and 400164 were made consecutively
and so appear as a single line near 2001.2. The 1996-1997 flux history
appears very similar to 1998 with the source remaining quite stable
within a factor of \aprx2.}
\end{figure}

\section{Observations} 

We observed \Grs\ four times with \chandra.  Table~\ref{t:obs} lists the
observation dates and durations. Two HRC-I observations were made in 2000
September-October spanning an intermediate-to-hard spectral transition
\citep[identified with \rxte, ][]{Smi01a,Smi01b}.  
Another HRC-I and an ACIS/HETG observation were made back-to-back in
2001 March following a dramatic hard-to-soft transition to a
\emph{very} low flux state \citep{Smi01b}.  
Figure~\ref{f:lc} shows the \rxte/PCA light curve with our \chandra\
observations indicated.  In each observation, a highly
significant point source was detected at a location consistent with
the known \Grs\ position.  In this paper we discuss the accurate
\xray\ position and morphology of \Grs\ as observed with the HRC and
in the HETG zero order image.  Preliminary results, both spatial and
spectral, from these observations were reported in \citet{Hei01}.
\begin{table}

\caption{\label{t:obs}Observations}
\begin{center}
\begin{minipage}{1.2\textwidth}
\begin{tabular}{lcccc} \hline\hline
        &	        &	 & Exposure	&  Roll Angle	\\
Seq. \#	&  Date		&  Inst. &(ksec)	&  Degrees   	\\ \hline
400085	& 2000 Sep 11.2	&  HRC-I & 1		&  270.5	\\
400131	& 2000 Oct 27.4	&  HRC-I & 10		&  268.5	\\
400164	& 2001 Mar 24.3	&  HRC-I & 10		&  89.8		\\ 
400163	& 2001 Mar 24.4	&  ACIS-S\footnote{HETG order zero.} & 30 & 90.1\\\hline\hline
\end{tabular}
\end{minipage}
\end{center}
\end{table}

\section{Analysis and Results}
\label{s:anal}

The HRC observations were processed with ASCDS version R4CU5UPD14.1,
and the ACIS-S observation with version R4CU5UPD14.5.  All four
datasets were processed with CALDB version 2.4. This is particularly
important for the HRC data, as CALDB versions 2.2 and later
incorporate up to date ``degapping'' coefficients necessary to provide
artifact free images \citep{Kenter01}.  For this work, we applied the
absolute pointing corrections available as of 2002 May 2\\
({\tt http://cxc.harvard.edu/cal/ASPECT/celmon}) \\ which accounts for slight
differences with the results reported in \citet{Hei01}.  We used
CIAO~2.2 (\chandra\ Interactive Analysis of Observations) tools
throughout these analyses.
\begin{figure}
%
\plotone{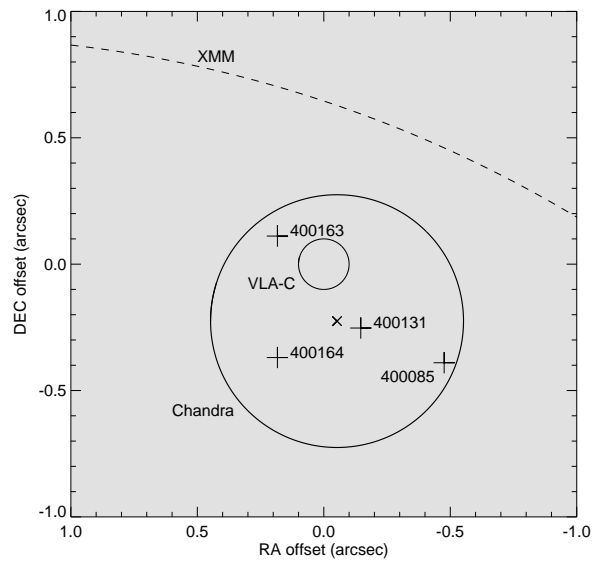}
\caption{\label{f:posn}Error circles for \Grs. The 90\% confidence \chandra\ error
circle (0\asec.45 radius) includes the 0\asec.1\ VLA-C radio position
of \citep{Mar98}. Positions from the four individual observations are
indicated by crosses and labeled by sequence number. Coordinates are
offsets from the radio position: (J2000) RA $=$ 18 01 12.395, Dec $=$
$-$25 44 35.90.  The 5\asec\ error circle from \emph{XMM} is also
indicated \citep{Gol01}.}
\end{figure}

\subsection{Source Location}
\label{s:posn}
To determine the position of \Grs, we extracted an image of the
central region of the field for each observation, binned at the full
level~2 events file resolution (\aprx0\asec.13 per pixel for the HRC
and \aprx0\asec.49 for ACIS-S).  We then applied the CIAO tool ``{\tt
wavdetect}'' to find the coordinates of the source.  The uncertainty
in the location of the source within the image is small compared to
the overall \chandra\ aspect uncertainties. According to \chandra\
\xray\ Observatory Center documents (see {\tt 
http://asc.harvard.edu/cal/ASPECT}), the RMS uncertainty in absolute
position determinations at the time of our observations was 0\asec.6,
corresponding to a 90\% confidence radius of $\rm R_{90} =$ 0\asec.9
for a single pointing.  The measured deviation of our four positions
is 0\asec.41~RMS, consistent with these calibrations. We averaged the
four measurements  (applying equal weights to each and assuming the
advertised 0\asec.6 RMS uncertainty) to derive $\mathbf{ R_{90} =
0\asec.45}$.
This assumes that the positions are Gaussian distributed and that no
systematic error causes an offset in the resulting mean position. We
note that two observations were made with spacecraft roll angles near
90\degree (see Table~\ref{t:obs}), while the others had roll angles
near 270\degree.  This would tend to cancel any roll angle dependent
systematic offsets.

Figure~\ref{f:posn} shows the measured source locations
from our four \chandra\ observations as well as the estimated
confidence region.  The ACIS-S/HETG position (400163) was based on the
zeroth-order image. While the image was significantly piled up, we
expect a negligible effect on the position \citep{Marsh01}.  The
\chandra\ 0\asec.45 radius error circle is centered at (J2000)
\emph{\bf{ RA $=$
\grsra, Dec $=$ \grsdec}}.

We note that a second source (designated \\ CXOU~J180106.1$-$253903; RA
$=$ \alphara, Dec $=$ \alphadec, $\rm R_{90} = 0\asec.9$) was
detected in all three HRC pointings, and a third
(CXOU~J180031.0$-$254154; RA $=$ \betara, Dec $=$ \betadec,
$\rm R_{90} = 1\asec.7$) was marginally detected in a single pointing
(see Table \ref{t:rates}).  The \chandra\ position
for CXOU~J180106.1$-$253903 is 0.76\asec\ from a USNO-A2 star with red
and blue magnitudes of 14.7 and 15.9 respectively.  While this star
may be the optical counterpart of CXOU~J180106.1$-$253903, no
astrometric reference lies within 5\asec\ of CXOU~J180031.0$-$254154.
A single precision frame tie, particularly 6\amin\ off axis, is
inadequate to improve dead-reckoning astrometry. Therefore, we have
not applied any aspect correction based on the possible identification
of CXOU~J180106.1$-$253903.

\subsection{Image Analysis}
\label{s:morph}

\begin{figure}
%
\plotone{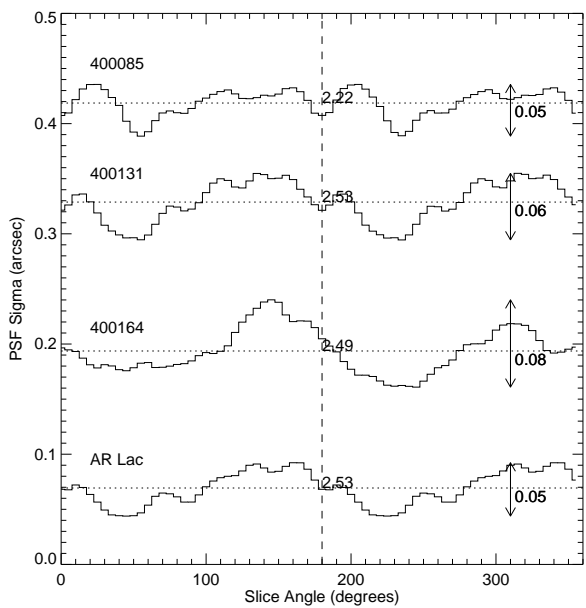}
\caption{\label{f:psf} Sigmas of Gaussian fits to slices through the
image peak as a function of azimuthal angle for the three HRC
observations of \Grs\ and, for comparison, AR~Lac.  For clarity, each
observation has been offset and the mean value of sigma labeled and
indicated by a dotted line.  The amplitude of the variations are
indicated at the right. }
\end{figure}

To search for extended emission (e.g. jets) which would appear as
asymmetries in the image peak, we performed Gaussian fits to slices
through the peak taken over a range of position angles.  We performed
the same analysis on an HRC observation of the RS CVn binary,
AR~Lac. AR~Lac is expected to be a point source, and given its very
low column density \citep[$N_H < 2\times10^{18}$\,\percmsq ][]{Kaa96},
should also lack any significant scattering halo.  Figure~\ref{f:psf}
shows the resulting Gaussian sigmas plotted as a function of slice
position angle in detector coordinates for all three \Grs\ and the
AR~Lac HRC observations.  In all cases, a roughly sinusoidal variation
is seen with an amplitude of \aprx0\asec.06.  The phase of the
variation is the same in all observations.  We note that observation
400164 has a slightly higher amplitude (0\asec.08), but because the
phase is unchanged and the overall size of the effect is so small, we
conclude that this is not indicative of source structure. We
therefore attribute these variations to instrumental effects, and
conclude that \Grs\ is consistent with a point source.

\section{Discussion}

\citet{Gol01} recently used \emph{XMM} data to derive a 5\asec\ error
radius for \Grs. This work, which is consistent with their result,
reduces the area of the error region by two orders of magnitude. 
The coincidence of the sub-arcsecond VLA and \chandra\ error circles
seals the association of \Grs, the \xray\ source, with the variable
radio source \citep[``VLA-C'', ][]{Mar98}. From radio source counts
\citep{Condon84}, we estimate a chance probability of \aprx$10^{-5}$ for
a radio source brighter than VLA-C (0.2\,mJy) to fall in the \chandra\
error circle.  \citet{Mar98} identified two candidate counterparts to
VLA-C in I and K band images. The brighter and closer candidate was
found through multi-band photometry and near infrared spectroscopy
most likely to be an early K giant.  Revised astrometry \citep{Roth02}
of infrared observations by \citet{Eik01} confirm that this star (labeled `A') is
consistent with VLA-C at the 3$\sigma$ level.  The second candidate of
\citet{Mar98} is likely a main sequence F star, but is inconsistent
with the astrometry of \citet{Roth02}.  Furthermore, \citet{Smi02}
have found a low amplitude (\aprx4\%), $18.45\pm 0.10$\,dy periodic
modulation of the \Grs\ \xray\ emission observed with \rxte.  This
period fits well with the orbit of a Roche lobe filling K giant
companion \citep{Roth02}, and so a more complete picture of \Grs\ is
emerging.  With the \xray/radio association confirmed and periodic
\xray\ emission pointing to a K giant, \Grs\ appears to be a black
hole with an intermediate mass giant companion.  

GRS~1915$+$105 \citep{Gre01}, GRO~J1655$-$40 \citep{Shahbaz99},
XTE~J1550$-$564 \citep{Orosz02}, and now
\Grs\ -- all four of the black hole microquasars in low mass systems
-- have evolved companions.  Meanwhile, jets have not been observed in
the more common black hole soft \xray\ transients (SXTs) with main
sequence companions.  This prompts us to examine the connection
between giant companions and jet activity. Perhaps an evolved
companion overflows its Roche lobe by a large factor, providing a
higher typical accretion rate. These four objects are persistent or
quasi-persistent, having very different time histories from the
standard SXT ``fast rise and exponential decay'' light curve.  The SXT
outbursts are driven by an accretion instability following long
periods of relatively low accretion rates \citep[see for example,
][]{Cannizzo98}.  Possibly steady, high rate accretion provides an
environment conducive to jet formation.

Finally, we note that the present lack of \xray\ jets is completely
consistent with the energetics of the radio lobes. \citet{Ro92},
assuming a distance of 8.5\,kpc, estimated the luminosity of the radio
lobes to be \aprx$10^{30}$\,\lumin and their energy content to be roughly
$10^{44}$\,ergs.  With a typical \xray\ luminosity of a few$\times
10^{37}$\,\lumin, the present lobes could have been energized by a
small fraction of the \xray\ power over a period as short as a few
years. In fact, given the long lifetime (\aprx$3\times10^{6}$\,yr) of the
radio lobes, this indicates that either the duty cycle for or
efficiency of feeding the jets (or both) is very low.

\acknowledgements

This work was supported by \chandra\ General Observer Project awards
GO0-1116A and GO1-2035X.






\begin{table}
\caption{\label{t:rates}Counting rates}
\begin{center}
\begin{minipage}{1.2\textwidth}
\begin{tabular}{lccc} \hline\hline
        & \multicolumn{3}{c}{Rates (counts/s)} \\ \cline{2-4}
Seq. \#	& \Grs 	            &CXOU~J180106.1$-$253903        & CXOU~J180031.0$-$254154 \\ \hline
400085	&  $11.27 \pm 0.01$ & $(7.0 \pm 3.1)\times 10 ^{-3} $ & -- \\
400131	&  $4.18 \pm 0.02$  & $(5.3 \pm 0.9)\times 10 ^{-3} $	&   $(6.5 \pm 1.3)\times 10 ^{-3} $ \\
400164	&  $7.68 \pm 0.03$  & $(3.3 \pm 0.8)\times 10 ^{-3} $	&  --  \\ 
400163\footnote{ACIS/HETG order zero.}	&  $0.130 \pm 0.004$ &  -- & --  \\\hline\hline
\end{tabular}
\end{minipage}
\end{center}
\end{table}

\newpage


\end{document}